\documentclass[letterpaper, 10 pt, conference]{ieeeconf}  

\IEEEoverridecommandlockouts                              
\usepackage{hyperref}
\usepackage{graphicx}
\graphicspath{ {} }

\title{{\LARGE \bf
The Salpeter IMF and its descendants}\\
{\large 100 YEARS OF THE IAU - Nature Astronomy news and views}\\
{\small VOL 3 | JUNE 2019 | 482-484, DOI: 10.1038/s41550-019-0793-0}
}

\author{ \parbox{3.4 in}{\centering Pavel Kroupa\\
         {\footnotesize Helmholtz-Institut f\"ur Strahlen und Kern-Physik, Bonn, Germany}\\ 
         {\footnotesize Astronomical Institute, Charles University in Prague, Czech Republic} \\
         {\tt\small pkroupa@uni-bonn.de}\\
         \vspace{0.5cm}
         }
         \hspace*{ 0.3 in}
         \parbox{3.4 in}{ \centering Tereza Jerabkova\\
        {\footnotesize European Southern Observatory,Garching, Germany}\\
         {\footnotesize Helmholtz-Institut f\"ur Strahlen und Kern-Physik, Bonn, Germany}\\ 
         {\footnotesize Astronomical Institute, Charles University in Prague, Czech Republic} \\
         {\tt\small tjerabkova@gmail.com}}
}

\begin{document}

\maketitle
\thispagestyle{empty}
\pagestyle{empty}

\begin{abstract}
The stellar initial mass function (IMF) is the key to understanding the matter cycle in the Universe. Edwin Salpeter’s
paper of 1955 founded this research field. Evidence today, however, challenges the initial mass function as an
invariant probability distribution function.
\end{abstract}

\section*{}
By the mid-1950s, what powers stars became increasingly understood through the framework of the new theory of quantum physics, by then widely accepted. In 1951, Edwin Salpeter, born
to a Jewish family in Austria and having emigrated to Australia in 1939, published with Hans Bethe the Bethe-Salpeter equation\cite{c1}, which describes the bound states of two relativistic particles in quantum theory, and in 1952 he wrote a paper on how the triple-alpha-process could operate
to produce carbon through the beryllium-8 resonance\cite{c2}. With his 1955 landmark paper\cite{c3}, written at the Australian National University in Canberra while on leave from Cornell, Edwin Salpeter built a first bridge connecting the new theory of quantum mechanics to cosmology, therewith addressing the cosmic matter cycle and chemical enrichment over time. He showed how quantum mechanical processes
transpire to define the macroscopic phenomenon of stellar birth and death rates, given the astronomical observations of the visible stellar population in the Galaxy. From today’s vantage point, it is thus right to recognize his 1955 paper\cite{c3} to be one of the most important research papers
ever written in astrophysics. 

The understanding of the Galaxy and its stars as being one of many extragalactic nebulae was also in transformation, and the first evidence for the unexpected shape
of the rotation curves had emerged with the work of Babcock\cite{c4} on Andromeda. The
mass distribution in these nebulae was thus becoming a research problem. The realization that stars are born, evolve at a rate depending on their mass through fusing 12\% of their mass from hydrogen to helium, and then evolve off the main sequence to die, led to the possibility that the mass
distribution in a galaxy might change with time. Salpeter assumed that the stars do not
change their masses except at death, and that they are born and die at a constant rate. For the age of our Galactic disk, he adopted $T_0 = 6$ Gyr. These assumptions allowed calculation, for the first time,
of the “relative probability for the creation of stars of mass” $m$, $\xi(m)$, which today is
called the stellar initial mass function. 

To achieve this calculation, he used the total observed distribution of their luminosities reported as star counts. He very roughly corrected this total
luminosity function, $\Phi_t(M_V)$, where $M_V$ is the absolute magnitude of a star in the
visual photometric band, for contamination by non-main-sequence stars and white
dwarfs, yielding $\Phi(M_V)$. Edwin Salpeter noted the steepening of $\Phi(M_V)$ at
magnitudes brighter than $M_V \approx 5$. Using the observation that population II stars (which
he assumed to have an age near 6 Gyr) have a main sequence turn-off near
$M_V = 3.5$, he attributed the steepening to the stars with this magnitude having
lifetimes corresponding to the age of the Galactic disk. Salpeter accounted for the
stars already evolved off the main sequence by assuming that stellar death occurs
when 12\% of the hydrogen content is used up, that the lifetime of a star of $M_V = 3.5$
is $T_0$ = 6 Gyr and that the rate with which hydrogen is consumed is proportional to
$m/L$, where $L$ is the bolometric luminosity of the star. The lifetime of any star brighter
than $M_V = 3.5$ thus becomes $T_0(L_{3.5}/m_{3.5})(m/L)$, where $m_{3.5}$ and $L_{3.5}$ are, respectively, the mass and bolometric luminosity of a turn-off ($M_V = 3.5$) star of 6 Gyr age.
Thus, he derived the “original luminosity function”, $\Psi(M_V)$ (Fig. 1), which gives the
relative frequency of stellar luminosities if the stars did not evolve. 

As additional hypotheses, he invoked $\xi(m)$ to be independent of time and to be a smooth function of $m$. A key function
is the derivative of the stellar mass-luminosity relation, $\mathrm{d}(\log_{10}{m})/\mathrm{d}M_V$,
which Edwin Salpeter had to approximate roughly by using “a compromise between various compilations” (Fig. 1). Defining $\mathrm{d}N = \xi(m)\mathrm{d}(\log_{10}{m})$  to be the number of
stars in the logarithmic mass interval $\log_{10}{(m)}$ to $\log_{10}{(m)}$+$\mathrm{d}(\log_{10}(m))$, the calculation he performed yielded $\xi(m) \approx 0.03(m/M_{\odot})^{-1.35}$, which today is
known as the Salpeter IMF with power-law index $x = 1.35$, valid for $0.4 < m/M_{\odot} < 10$.
This IMF, when extrapolated and integrated over all stellar masses (from 0.3 $M_{\odot}$ to infinity), implied that the number of then known white dwarfs matched about 10\% of stars today on the main sequence, in agreement with the star counts. He also calculated the mass budget, showing that
the total mass in main-sequence stars today approximately compares with all mass ever
formed in evolved stars. From the matching amount of interstellar gas observed, he
suggested that most of it must have been recycled through stars, lending credence
to the original suggestion by Fred Hoyle that all elements apart from hydrogen were
fused in stars.

 It is interesting to consider which of Salpeter’s suggestions remain valid and how
the field has evolved. Zinnecker\cite{c5} noted that had Salpeter used the present-day
accepted value for the age of our Galaxy’s disk, $T_0 = 12$ Gyr, he would have obtained
$x = 1.05$, because, given the star counts, one would have had to add more massive
stars into the original luminosity function. Miller and Scalo\cite{c6} updated the Salpeter
IMF for a larger magnitude range, paying particular attention to uncertainties, while
also taking into account stellar brightening during main-sequence evolution and the
changing disk scale height with stellar age. They used improved stellar mass–luminosity
data and discussed the degeneracy between a time-variable star-formation rate and
the shape of the IMF (assumed constant). The authors found the stellar birth rate to
be largely constant over the age of the disk and the gas-depletion time to be 1–3 Gyr,
deducing that the Galaxy needs to be accreting gas to sustain its roughly constant
star-formation rate. The mass in all stars and stellar remnants was found to be
comparable to the gravitating mass in the disk of our Galaxy. A major advance in terms of depth
of discussion and taking all possible uncertainties and newer data into consideration came next in the monumental effort by Scalo\cite{c7}. Improved stellar-evolution theory and empirical stellar mass-luminosity data constraints for brighter stars and new constraints on $\Psi(M_V)$ were
used. The dip in $\Psi(M_V)$  near $M_V = 7$, previously dismissed, was now adopted as
a real feature (Fig. 1a). The breakthrough photographic-survey technology relying
on photometric parallax distances\cite{c8} allowed $\Psi_{\mathrm{phot}}(M_V)$ to be constrained much better at the faintest magnitudes, showing it to decrease significantly, in contrast to the
statistically much less constrained $\Psi_{\mathrm{trig}}(M_V)$ obtained from trigonometric parallax
distance measurements of the nearby stars (Fig. 1a). The IMF in ref. \cite{c7} yields a roughly
constant slope of $x \approx 1.7$ above m = 1 $M_{\odot}$, a dip near m = 0.65 $M_{\odot}$ and a decreasing IMF below $m \approx 0.2$ $M_{\odot}$ (Fig. 1b). The work
in ref. \cite{c7} thus forced the observed subtle features in $\Psi(M_V)$, noted but rejected by
Edwin Salpeter, into the IMF. The author also exhaustively discussed the IMF
constraints for star clusters and for pre-main-sequence stars, and extended the discussion to the implications of the IMF for the integrated light and chemical evolution of galaxies.

Further advances came with the work that included unresolved multiple stars,
Lutz–Kelker and Malmquist biases and pre-main-sequence dimming and main-sequence brightening\cite{c9}. The authors showed that the subtle features in $\Psi(M_V)$
are due to changes in the stellar mass-luminosity relation through the opacity, molecular weight and convective structure within late-type stars (Fig. 1c,d). By using $\Psi_{\mathrm{trig}}(M_V)$ and 
$\Psi_{\mathrm{phot}}(M_V)$ as the resolved single-star and unresolved multiple-star
luminosity functions, respectively, they solved for one $\xi(m)$. This smooth IMF, shown in Fig. 1b, predicts all stellar populations to have a sharp peak in $\Psi(M_V)$ near $M_V = 11.5$ (Fig. 1d) with a slight dependency on metallicity. An immense spectroscopic observational effort\cite{c10} of
different clusters and associations in various galactic environments established
x = 1.35 to be the correct value for the IMF of massive stars. The present-day generally accepted IMF\cite{c11} can thus be well represented by the canonical two-part
power-law function with x = 0.3 for
$0.1 < m/M_{\odot} < 0.5$ and x = 1.3 for $0.5 \leq m/M_{\odot} < 150$. This IMF shape
was confirmed\cite{c12}, but using only the trigonometric-parallax-based $\Psi_{\mathrm{trig}}(M_V)$.
The basic assumption made by Salpeter\cite{c3} that the star-formation rate of the Galactic
disk was roughly constant with evidence for some variation over a 1 Gyr timescale,
and that the field-star IMF has $x \approx 1.3$ for $m$
more than a few $M_{\odot}$, has been confirmed by modern star-count work\cite{c13}.

The present-day discussion has shifted to testing the original assumption by
Edwin Salpeter\cite{c3} that the IMF is a probability density function, or whether it might
rather be an optimal (highly self-regulated) distribution function\cite{c14}. Recent observations
have uncovered a dependency of the IMF on metallicity and density\cite{c15,c16}, and infrared
surveys have shown stars to form in long and thin molecular cloud filaments\cite{c17},
challenging the gravo-turbulent theory of the IMF. Returning to the caveat expressed
in ref. \cite{c3} that the deduced IMF is but an average of the star-formation activity over
the age of the Galactic disk: if the field IMF is constructed from individual star formation
events in a galaxy, this composite field or galaxy-wide IMF becomes time and
spatially variable \cite{c18}. This variation is consistent with the observationally deduced
variation of the galaxy-wide IMF in dwarf \cite{c19} and major \cite{c20} disk galaxies, as well as with the
seminal deduction that the galaxy-wide IMF is likely to have been top-heavy during the formation of elliptical galaxies \cite{c21}. The fact that the Scalo IMF is steeper than
the Salpeter–Massey index $x \approx 1.3$ above m = 1 $M_{\odot}$ (upper green IMF in Fig. 1b) can
be accounted for through the field-IMF being the result of the composition of many IMFs.

The modern discussion distinguishes between the stellar IMF as stemming from  the star-formation process in a molecular cloud core (a star-formation event or embedded cluster) and the galaxy-wide
IMF as being the composition of all such star-formation events. This leads to major implications for understanding galaxy formation and evolution and thus for cosmology\cite{c14}. It will be important to carry out further observational tests of the optimal nature of the IMF, its emergence from the
fragmentation of molecular cloud filaments and the role of gravo-turbulence, and to
probe extreme star-forming systems at high redshift to better constrain the variation
of the IMF \cite{c22}.

\addtolength{\textheight}{-12cm}   



{\it {\bf Acknowledgements --} We acknowledge discussions in the International Space Science Institute (ISSI/ISSI-BJ) in Bern and Beijing made possible through the funding of the team “Chemical
abundances in the ISM: the litmus test of stellar IMF
variations in galaxies across cosmic time” (Principal
Investigators D. Romano and Z.-Y. Zhang). }


\begin{figure*}[h]
  \includegraphics[width=18cm]{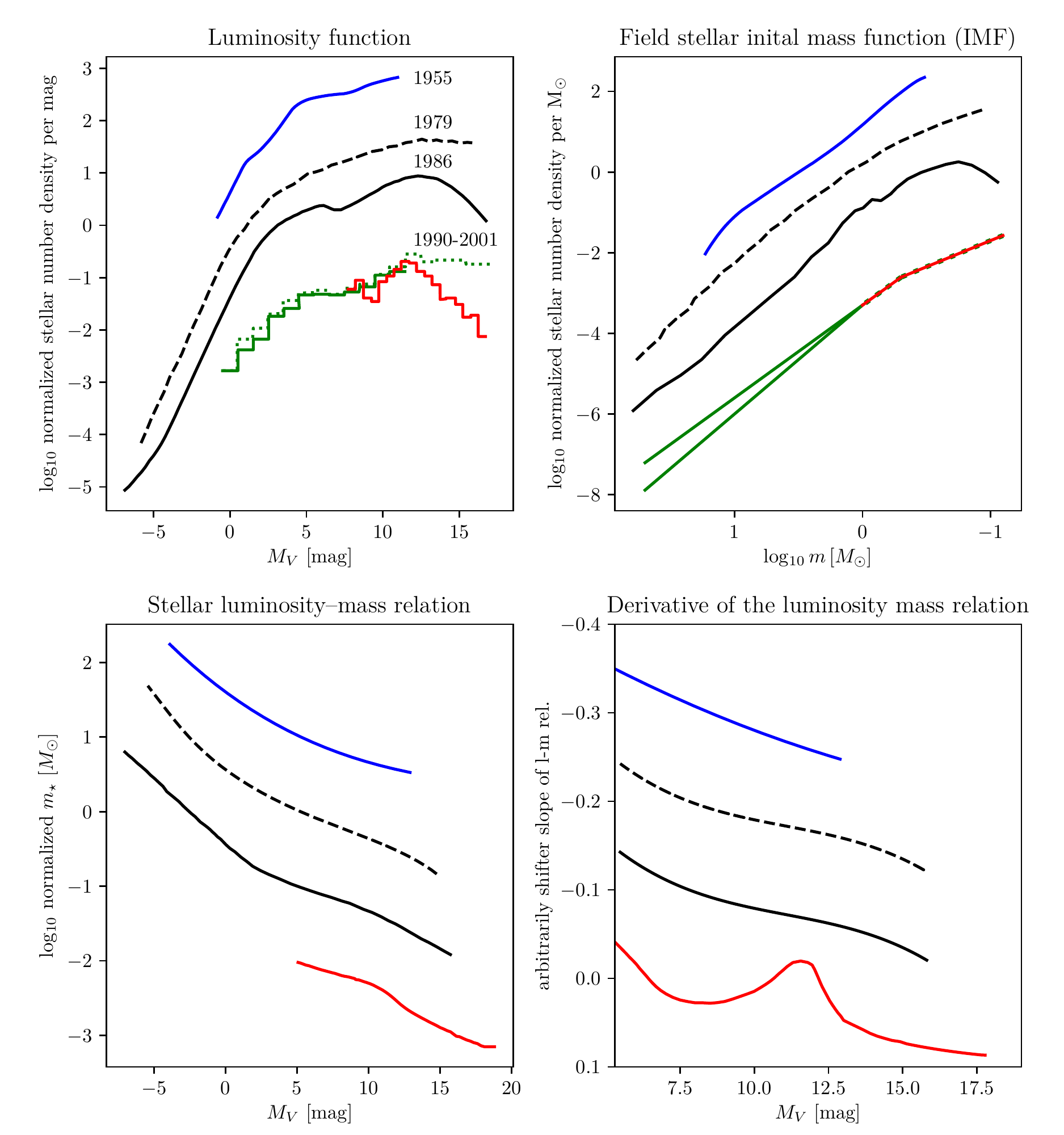}
  \caption{Different IMF-based properties for various IMFs: Salpeter IMF \cite{c1} (blue), Miller \& Scalo IMF \cite{c6} (dashed black), Scalo IMF \cite{c7} (solid black) and Kroupa IMF \cite{c9,c11} (green and red). {\bf a,} Luminosity function,
  $\Psi(M_V)$. $\Psi_{\mathrm{trig}}(M_V)$, dotted green for ground-based and solid green for space-based parallaxes; $\Psi_{\mathrm{phot}}$, solid red histogram. 
  {\bf b,} IMF, $\xi(m)$, red for $m<1 M_{\odot}$ and green for $m> 1 M_{\odot}$, 
  with the Scalo IMF being the lower line and the Salpeter-Massey IMF the upper line. 
  {\bf c,d,} Stellar mass-luminosity relation {\bf (c)} and its derivative {\bf (d)}. All curves are shifted vertically arbitrarily to demonstrate their differences.}
  \label{fig:graph}
\end{figure*}


\begin{thebibliography}{99}

\bibitem{c1} Salpeter, E. E. \& Bethe, H. A. Phys. Rev. 84, 1232–1242 (1951).
\bibitem{c2} Salpeter, E. E. Astrophys. J. 115, 326–328 (1952).
\bibitem{c3} Salpeter, E. E. Astrophys. J. 121, 161–167 (1955).
\bibitem{c4} Babcock, H. W. Lick Obs. Bull. 19, 41–51 (1939).
\bibitem{c5} Zinnecker, H. in UP2010: Have Observations Revealed Variable Upper End of the Initial Mass Function? (eds Treyer, M. et al.) 3–10 (Astronomical Society of the Pacific, 2011).
\bibitem{c6} Miller, G. E. \& Scalo, J. M. Astrophys. J. 216, 548–559 (1979).
\bibitem{c7} Scalo, J. M. Fund. Cosmic Phys. 11, 1–278 (1986).
\bibitem{c8} Reid, N. \& Gilmore, G. Mon. Not. R. Astron. Soc. 201, 73–94 (1982).
\bibitem{c9} Kroupa, P., Tout, C. A. \& Gilmore, G. Mon. Not. R. Astron. Soc.262, 545–587 (1993).
\bibitem{c10} Massey, P. Annu. Rev. Astron. Astrophys. 41, 15–56 (2003).
\bibitem{c11} Kroupa, P. Mon. Not. R. Astron. Soc. 322, 231–246 (2001).
\bibitem{c12} Chabrier, G. Publ. Astron. Soc. Austr. 115, 763–795 (2003).
\bibitem{c13} Mor, R., Robin, A. C., Figueras, F., Roca-Fabregas, S. \& Luri, X.Astron. Astrophys. Lett. 624, L1 (2019).
\bibitem{c14} Kroupa, P. et al. in Planets, Stars and Stellar Systems Vol. 5 (eds Oswalt, T. D. \& Gilmore, G.) 115–242 (Springer, 2013).
\bibitem{c15} Marks, M., Kroupa, P., Dabringhausen, J. \& Pawlowski, M. Mon. Not. R. Astron. Soc. 422, 2246–2254 (2012).
\bibitem{c16} Zhang, Z.-Y., Romano, D., Ivison, R. J., Papadopoulos, P. P. \& Matteucci, F. Nature 558, 260–263 (2018).
\bibitem{c17} Andre, P. et al. Astron. Astrophys. 518, L102 (2010).
\bibitem{c18} Jerabkova, T. et al. Astron. Astrophys. 620, 39–44 (2018).
\bibitem{c19} Lee, J. C. et al. Astrophys. J. 706, 599–613 (2009).
\bibitem{c20} Gunawardhana, M. L. P. et al. Mon. Not. R. Astron. Soc. 415,1647–1662 (2011).
\bibitem{c21} Matteucci, F. Astron. Astrophys. 288, 57–64 (1994).
\bibitem{c22} Jerabkova, T., Kroupa, P., Dabringhausen, J., Hilker, M. \& Bekki, K. Astron. Astrophys. 608, A53 (2017).


\end{thebibliography}
\end{document}